\documentclass[a4paper,12pt]{article}
\usepackage{epsfig}
\usepackage{graphicx}
\usepackage{amsmath}

\author{Karina I. Mazzitello$^{a}$ and Juli\'an Candia$^{b}$\\{}\\
$^a${\small\it CONICET and Departamento de F\'{\i}sica, Facultad de Ingenier\'{\i}a,}\\  
{\small\it Universidad Nacional de Mar del Plata, Mar del Plata, Argentina}\\
{\small Email address: kmazzite@mdp.edu.ar}\\
$^b${\small\it CONICET and Instituto de F\'{\i}sica de L\'{\i}quidos y Sistemas 
Biol\'ogicos,}\\{\small\it Universidad Nacional de La Plata, La Plata, Argentina}\\
{\small Email address: juliancandia@gmail.com}}

\title{Diffusion-Based Density-Equalizing Maps: an Interdisciplinary Approach to 
Visualizing Homicide Rates and Other Georeferenced Statistical Data}

\begin{document}
\maketitle

\begin{abstract}
In every country, public and private agencies allocate extensive funding to 
collect large-scale statistical data, which in turn are studied and analyzed in order to determine local, regional, national, and international policies regarding all aspects relevant to the welfare of society. 
One important aspect of that process is the visualization of statistical data with embedded geographical information, which most often relies on archaic methods such as maps colored according to graded scales. In this work, we apply non-standard visualization techniques based on physical 
principles. We illustrate the method with recent statistics on homicide rates in Brazil and their correlation to other publicly available data. 
This physics-based approach provides a novel tool that can be used by interdisciplinary teams investigating statistics and model projections in a variety of fields such as economics and gross domestic product research, public health and epidemiology, socio-demographics, political science, business and marketing, and many others. 
\end{abstract}

\section{Introduction}
Nowadays, interdisciplinary efforts in science directed towards the most acute real-world problems of society are being increasingly encouraged and stimulated. And rightly so, this task is even more urgent in the Latin American region, which suffers from some of the highest rates of 
crime and violence in the world~\cite{flacso} 
and their devastating effects on citizen security and human rights~\cite{oea}, as well as from other endemic 
issues such as widespread illiteracy, sharp socioeconomic disparities, undernourishment, etc. 
In this context, the statistical physics community is in an excellent position to 
contribute invaluable new tools and insight to areas in which statistical analysis is performed by long-standing, more traditional means.
 
One of the important aspects of dealing with statistical information is data visualization. 
Indeed, as we enter the age of the petabyte, where we routinely capture, warehouse and analyze massive amounts of data~\cite{petaage}, 
pioneering methods of visualization are needed to change the way we understand and disseminate science.
Some examples of new visualization techniques recently developed are state-of-the-art animations to see the quantum world, 
accurate maps of the universe based on the Sloan Digital Sky Survey covering large areas of the sky, and complex network representations that facilitate the recognition of patterns in multidimensional complex systems~\cite{physicsworld}. 

In the field of applied statistics, traditional visualization tools such as bar and pie charts, regression plots, and box plots, are still 
the most widely used methods. Statistical data with embedded geographical information (for instance, census data) are either organized in tables 
or shown by means of standard maps with color codes by region. Such maps, however, can be misleading. Often times, statistical 
measures are correlated with the population density and other indicators, but those correlations remain hidden from the representation. 
For instance, a plot of disease incidence will inevitably show high incidence in cities and low incidence in rural areas, solely because 
more people live in cities. An alternative representation is to plot the incidence per capita, which solves this problem at the cost 
of discarding all information about where most of the cases are occurring. In order to 
meaningfully visualize statistical data, bringing together relevant statistical indicators and the embedded geographical information, a 
different approach is needed. 

Over the last few decades, some visualization work has focused on creating so-called {\it cartograms}, 
in which geographic regions are deformed and rescaled 
in proportion to their population or other statistical indicators. Several of these efforts were inspired by physics. For instance, 
in the approach by Gusein-Zade and Tikunov~\cite{guse93}, a continuous displacement field is used to expand high population areas by 
means of a repulsive force. Other methods by Appel et al.~\cite{appe83} and by Dorling~\cite{dorl96} are based on cellular automata, 
in which local interaction rules allow boundaries to shrink or expand locally until equilibrium is reached. Another procedure proposed by 
Kocmoud~\cite{kocm97} relies on a mechanical model of masses and springs, which is constained in order to maintain some topographic features. 
Despite these efforts, cartograms built following those methods proved problematic in a number of ways, leading to highly distorted 
shapes, overlapping regions, or strong dependences on the choice of coordinate axes. 
More recently, Gastner and Newman~\cite{gast04} proposed a new method for producing cartograms based on diffusion equations, which 
resolves the shortcomings experienced with previous techniques. Starting with an inhomogeneously distributed statistical quantity (e.g. 
the population density), diffusion equations are numerically integrated with appropriate boundary conditions. By allowing the diffusion 
process to evolve until an homogeneous equilibrium state is reached, the displacements can be reinterpreted to generate a cartogram. 

Within this context, 
in this work we implement the diffusion-based method to create a variety of density-equalizing 
maps of the 27 Brazilian federative units. By using this approach,  
we generate a series of cartograms that show different ways of displaying population, homicide rates, and a combined index of 
education, health, and economic development. Depending on the way the cartograms are built and how the statistics are chosen, 
different perspectives can be gained from the data. Also, cross-correlations between statistical quantities and their dynamical 
evolution can be represented in more effective ways than standard visual methods.  
The aim of this paper is to show how a novel physics-based 
visualization tool can be applied to real-world data of great cross-disciplinary impact, as well as to discuss the benefits that 
can be drawn from the new approach compared to the visual methods usually employed 
by statisticians of public and private agencies.    

The rest of this paper is organized as follows. In Section 2, we present details on the diffusion-based method to generate density-equalizing 
maps. Section 3 is devoted to the presentation and discussion of the cartograms, 
while our conclusions are finally stated in Section 4.  

\section{The Diffusion-Based Cartogram Method} 

In this Section, we summarize the diffusion-based method for generating density-equalizing maps~\cite{gast04}. 
Let us consider $\rho_0({\vec{r}})$ the spatial density distribution of the inhomogeneous statistical indicator that we want to represent by means of a rescaled map. 
We will call $\rho_0({\vec{r}})$ the {\it population density}, since the population distribution is one of the 
indicators one may be interested in. Notice, however, that the method can be applied to any other statistical quantity distributed inhomogeneously in space, as will be shown in the examples of Section 3. The basic idea is to rescale the map's regions to become  proportional 
to their population, so that the population density on the rescaled map is homogeneously distributed. 
If the total area of the map is preserved, the homogeneous population density after rescaling will be equal to the mean 
population density before rescaling, $\bar{\rho_0}$.  

From a mathematical point of view, any total-area-preserving map rescaling operation corresponds to a coordinate transformation 
${\vec{r}}\to{{\vec{r}}'}$ under the condition 
that the determinant of the corresponding Jacobian matrix equal the density ratio $\rho_0({\vec{r}})/\bar{\rho_0}$.    
Although this condition guarantees that the total area of the map remains constant, 
it does not provide specific information about 
the cartogram projection. In order to determine a unique projection, we 
resort to the physical process of diffusion by considering the population density as a time-dependent scalar field, 
$\rho({\vec{r}},t)$, that satisfies the initial condition $\rho({\vec{r}}, t=0)\equiv\rho_0({\vec{r}})$. 
From a macroscopic perspective, the current density, related to time-dependent velocity and density fields through
\begin{equation}
{\vec{J}} = {\vec{v}}({\vec{r}},t)\rho({\vec{r}},t)\ ,
\end{equation}  
\noindent is directed along the gradient of the density field:
\begin{equation}
{\vec{J}} = -D \nabla \rho\ ,
\end{equation} 
\noindent where $D$ is the diffusion constant. Without loss of generality, we can conveniently select 
time and length units in order to set $D\equiv 1$. 
These equations, together with the equation for the conservation of mass,
\begin{equation}
\nabla\cdot {\vec{J}} + {\frac{\partial\rho}{\partial t}} = 0\ ,
\end{equation} 
lead to the well-known diffusion equation:
\begin{equation}
\nabla^2\rho-{\frac{\partial\rho}{\partial t}} = 0\ .
\end{equation} 
We enclose the area of interest within a larger box, whose initial population density matches the mean population density ${\bar{\rho_0}}$. 
The role of this box is merely to act as a large container and avoid finite-size effects due to the system's boundaries. 
By solving this diffusion equation with the initial condition 
$\rho({\vec{r}}, t=0)\equiv\rho_0({\vec{r}})$ and Neumann boundary conditions such that no flow occurs through the boundaries of the container, the density field can be determined at all times. Then, the velocity field 
follows as 
\begin{equation}
{\vec{v}}({\vec{r}},t)=-{\frac{\nabla\rho({\vec{r}},t)}{\rho({\vec{r}},t)}}
\end{equation} 
and from the latter, the cumulative displacement of any point in the map can be calculated by solving the kinematics, i.e. 
\begin{equation}
{\vec{r}(t)} = {\vec{r}}(0) + \int_0^t{\vec{v}}({\vec{r}},t')dt'\ .
\end{equation} 

The cartogram is thus derived by moving all region boundaries on the map in such a way that the net flow passing through them is zero. 
In the large-$t$ limit, the population is uniform everywhere within the box and the regions become rescaled to sizes 
proportional to their initial population densities.  

\begin{figure}[t!]
\centerline{{\epsfxsize=4.2in \epsfysize=2.5in \epsfbox{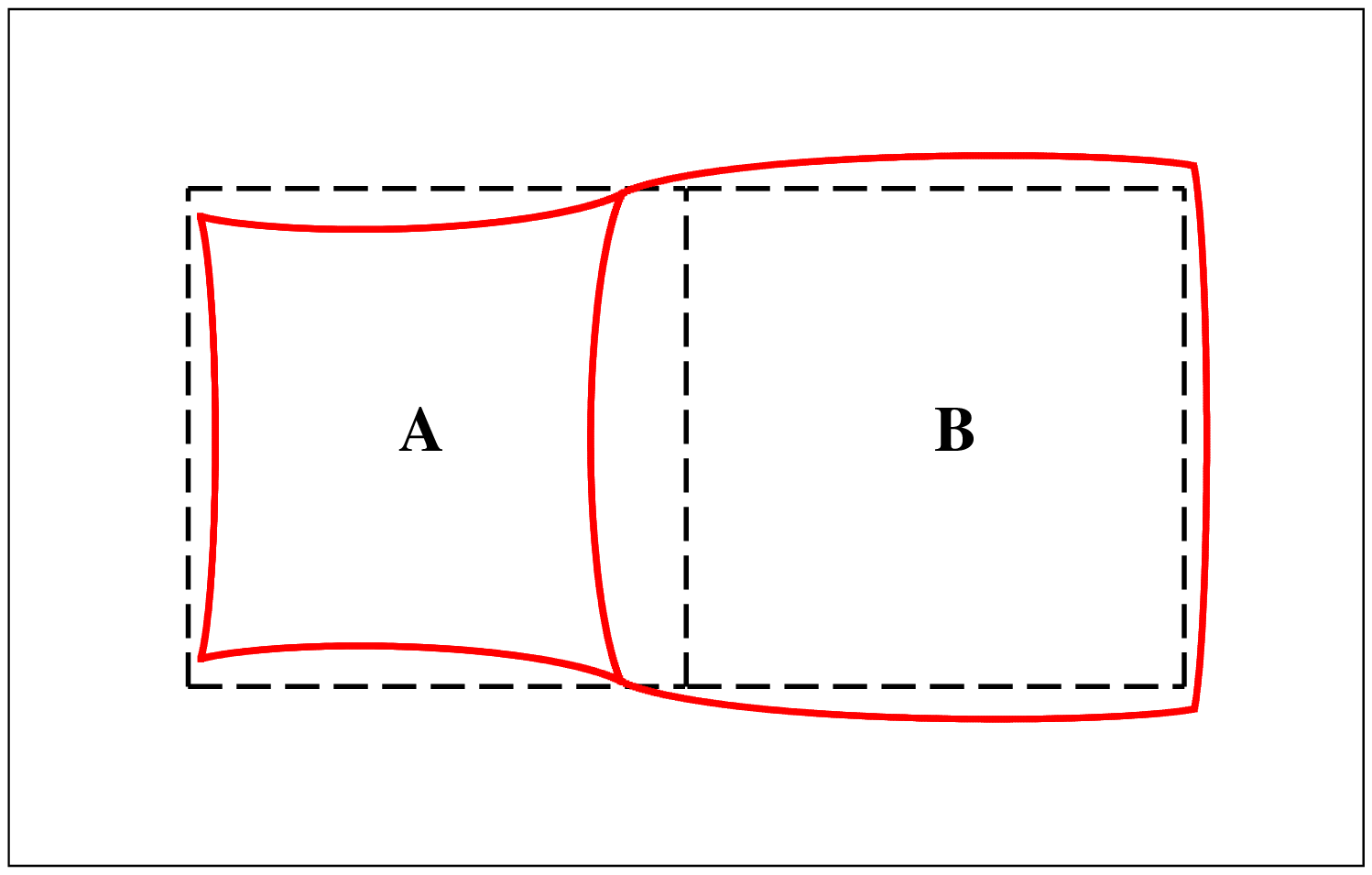}}}
\caption{Example showing the initial areas (enclosed by black dashed lines) and final areas (enclosed by red solid lines) 
when the population density of region B is twice as large as that of region A.}
\label{example}
\end{figure}

As a simple example to illustrate the procedure, consider the system shown in Fig.~\ref{example} 
by the black dashed lines. 
It consists of 2 regions of equal size: region A on the left 
has population density $\rho_{0,A}=1$ (in arbitrary units) and region B on the right has $\rho_{0,B}=2$. We embed both regions within a large enclosure 
filled with the average density $\bar{\rho_0}=1.5$. Letting the concentrations diffuse, the system evolves towards homogeneity. Once equilibrium 
is attained, the region boundaries have been reshaped as shown by the red solid lines. 
The rescaled map shows region B twice as large as 
region A, as expected from the initial population density ratio. That is, the ratio of final areas when the diffusion 
process reaches equilibrium, $Area(B)/Area(A)=2$, agrees with the ratio of initial densities, $\rho_{0,B}/\rho_{0,A}=2$.  
In this way, the rescaled map has merged the geographical information with the 
statistical information: while regions A and B are still recognizable, the rescaled representation also tells us 
that B is twice as large as A in the sense of the chosen statistical measure. The next Section is devoted to 
actual examples showing different ways to implement these ideas.   

\begin{figure}[t!]
\centerline{{\epsfxsize=3.5in \epsfysize=3.5in \epsfbox{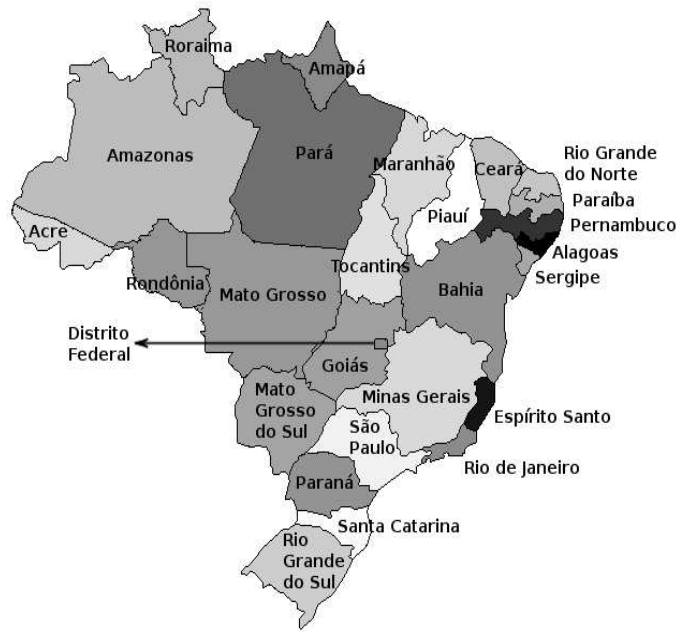}}}
\caption{Standard visualization of georeferenced statistical data: the shape and area of each 
Brazilian state is a faithful geographical representation, while the grayscale shading shows homicide rates per 100,000 
inhabitants using 2008 data from Ref.~\cite{wais11}. The highest homicide rate is 60.3 (Alagoas, in black) and the lowest one 
is 12.4 (Piau\'{\i}, in white).}
\label{usual_representation}
\end{figure}

\section{Density-Equalizing Cartograms of Brazil: Results and Discussion}

In this Section, we implement the diffusion-based method to generate density-equalizing cartograms of Brazil's states. By using recent, publicly released statistical data provided by Brazilian government agencies, we illustrate the method and discuss different scenarios to apply this visualization technique. 

Being one of the major challenges facing Brazil and, on 
a larger context, the rest of Latin American countries as well, we focus on 
the geographical distribution of homicide statistics, their cross-correlation with other 
statistical measures, and their time evolution. 
To that end, we examined data on nationwide violent crime statistics very recently released in a joint report by the Brazilian Ministry of Justice and the Instituto Sangari~\cite{wais11}. Demographic data was obtained from the IBGE~\cite{ibge}, 
Brazil's chief government agency responsible for statistical and geographical information, whose Portuguese acronym stands for 
``Brazilian Institute of Geography and Statistics".  
It is interesting to point out, in passing, that the tool presented in this work integrates {\it geography} and {\it statistics} into a unified visual framework, thus naturally reconciling the two main subject areas of this agency. Furthermore, we used the IFDM index from the so-called ``FIRJAN 
System"~\cite{firjan}, which measures the combined education, health, and economic development at the 
municipal level based on official data released by the Brazilian Ministries of Education, Health, and Labor.

\begin{figure}[t!]
\centerline{{\epsfxsize=3.5in \epsfysize=3.5in \epsfbox{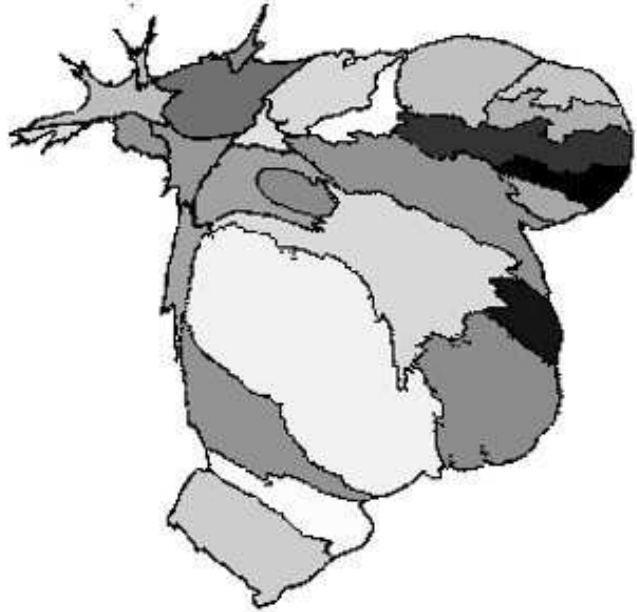}}}
\caption{Cartogram showing areas proportional to each state's population and grayscale shading corresponding to 
homicide rates per 100,000 inhabitants in 2008. Notice the shrinkage of the Northern and Center-West regions in constrast 
with the expansion of the Federal District and the 
Southern, South-Eastern, and North-Eastern regions. By paying the penalty of a highly 
distorted geography, we integrate the information of homicide rates (shading) with population size (areas).}
\label{absolute_popul_crime}
\end{figure}

Homicide rates by state are usually depicted as in Fig.~\ref{usual_representation}, in which the shapes and areas 
of the states are faithful geographical representations, while the different homicide rates are shown by means of 
color or grayscale shadings. This accurate geographical information comes at a price: from this representation, we 
do not know which states or regions are the ones where most homicides occur. Moreover, standing out in this map are 
the states with large surface areas, which might not be relevant neither by having particularly high homicide rates nor high absolute number of homicide counts. A good example in this regard is offered by the largest state, Amazonas, which 
ranks $17^{th}$ among the 27 federative units (i.e. the 26 states plus the Federal District) in the list of states with highest homicide rates, 
and also $17^{th}$ in the list of states with highest absolute homicide counts (using statistics corresponding to 2008~\cite{wais11}). In contrast, the top-ranking state in homicide rates, Alagoas, appears barely noticeable due to its 
very small surface area.     

\begin{figure}[t!]
\centerline{{\epsfxsize=3.5in \epsfysize=3.5in \epsfbox{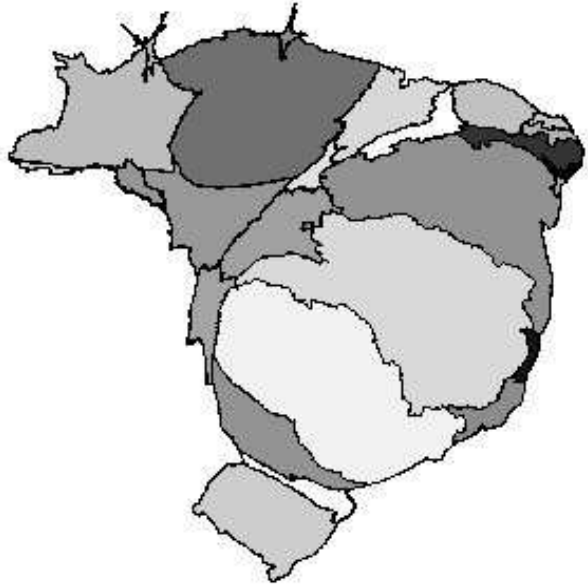}}}
\caption{Cartogram showing the change of areas (relative to its geographically faithful size) proportional to each state's population and grayscale shading corresponding to homicide rates per 100,000 inhabitants in 2008. This cartogram is a 
compromise between the standard representation of Fig.~\ref{usual_representation} and the highly distorted cartogram of 
Fig.~\ref{absolute_popul_crime}.}
\label{relative_popul_crime}
\end{figure}

In order to put in evidence the most populated states, we generate the cartogram of Fig.~\ref{absolute_popul_crime}  
by assuming $\rho_0({\vec{r}})\propto N_i/A_i$ $\forall {\vec{r}}\in A_i$, where $N_i$ is the population and $A_i$ 
the surface area of the $i-$th state. The density-equalizing diffusion-based method thus leads to a representation in which 
the area of each state is in proportion to its population. As in Fig.~\ref{usual_representation}, the grayscale 
shading shows each state's homicide rates per 100,000 inhabitants in 2008. Since the absolute number of homicides is 
obtained from the product of homicide rate times the population size, this representation arguably displays all three 
statistics (i.e. population, homicide rate, and homicide counts). 
For instance, large dark states in the North-Eastern region are emphasized, as well as Espirito Santo and Rio de 
Janeiro, as most affected by violent crime both in relative and absolute terms. Although having one of the lowest homicide 
rates in Brazil, S\~ao Paulo stands out at the center of the cartogram due to its being the most populated state 
in the country. Naturally, the cartogram approach trades off geographical accuracy in exchange for incorporating additional  
information. In the following figures, we discuss alternative ways to implement such trade-offs. 

Fig.~\ref{relative_popul_crime} shows a compromise between the geographically accurate, standard map of Fig.~\ref{usual_representation} and the highly distorted cartogram of Fig.~\ref{absolute_popul_crime}. In this cartogram, 
which is obtained from assuming $\rho_0({\vec{r}})\propto N_i$ $\forall {\vec{r}}\in A_i$, the {\it change} of area 
(relative to the geographically accurate representation) is proportional to each state's population. The grayscale shading, 
as before, corresponds to homicide rates per 100,000 inhabitants in 2008. Therefore, each state's area on this cartogram 
conveys some information on the geographical surface area, which was completely missing from the cartogram of Fig.~\ref{absolute_popul_crime}. Moreover, the shapes are less distorted than in the previous cartogram. It should be pointed out, 
however, that since this cartogram integrates geographical and statistical information in a nontrivial manner, its interpretation 
is rendered more difficult. For instance, the similarity of two states depicted with similar sizes on the cartogram could 
be due to different combinations of population and geographical surface area of those states. In Fig.~\ref{relative_popul_crime}, S\~ao Paulo and Minas Gerais appear represented with similar sizes, which is due to the fact that 
S\~ao Paulo's population is roughly twice as large, while its surface area is roughly half as large, as those of Minas Gerais.  

\begin{figure}[t!]
\centerline{{\epsfxsize=3.5in \epsfysize=3.5in \epsfbox{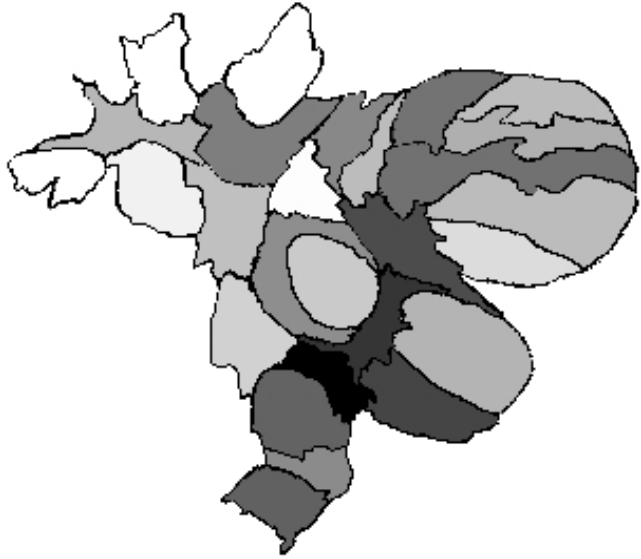}}}
\caption{Cartogram showing areas proportional to each state's homicide rates per 100,000 inhabitants in 2008 
and logarithmic grayscale shading corresponding to each state's population. In this case, the cartogram tends to enhance the 
appearance of states and regions with high crime rates, as opposed to those with high population (compare with 
Fig.~\ref{absolute_popul_crime}). For instance, Roraima and Amap\'a in the Northern region stand out in this 
cartogram due to their significant crime rates, but go unnoticed in Fig.~\ref{absolute_popul_crime} due to their 
small population.}
\label{absolute_crime_popul}
\end{figure}

In Fig.~\ref{absolute_crime_popul}, the areas are proportional to each state's homicide rates per 100,000 inhabitants in 2008, 
while the logarithmic grayscale shading reflects each state's population. 
That is, this figure reverses the roles of the two statistics 
shown earlier in Fig.~\ref{absolute_popul_crime}. This cartogram is obtained by assuming $\rho_0({\vec{r}})\propto R_i/A_i$ $\forall {\vec{r}}\in A_i$, where $R_i$ is the homicide rate of the $i-$th state. The cartograms in 
Figs.~\ref{absolute_popul_crime} and \ref{absolute_crime_popul} contain the same information, yet their appearance is 
very different. 
One could argue that the focus of attention is first guided towards the large states and, therefore, size takes precedence over 
shade. Under that assumption, Fig.~\ref{absolute_popul_crime} places more emphasis on population (by noticing first the most populated states and only later their corresponding homicide rates), while  
conversely, Fig.~\ref{absolute_crime_popul} emphasizes homicide rates. Although the preferential focus of attention (size, 
shape, or color) might depend on the subject, it is interesting to point out that, as exemplified here, the same data can be 
visualized in different ways that imply a different emphasis on the represented quantities. 

\begin{figure}[t!]
\centerline{{\epsfxsize=4.0in \epsfysize=3.5in \epsfbox{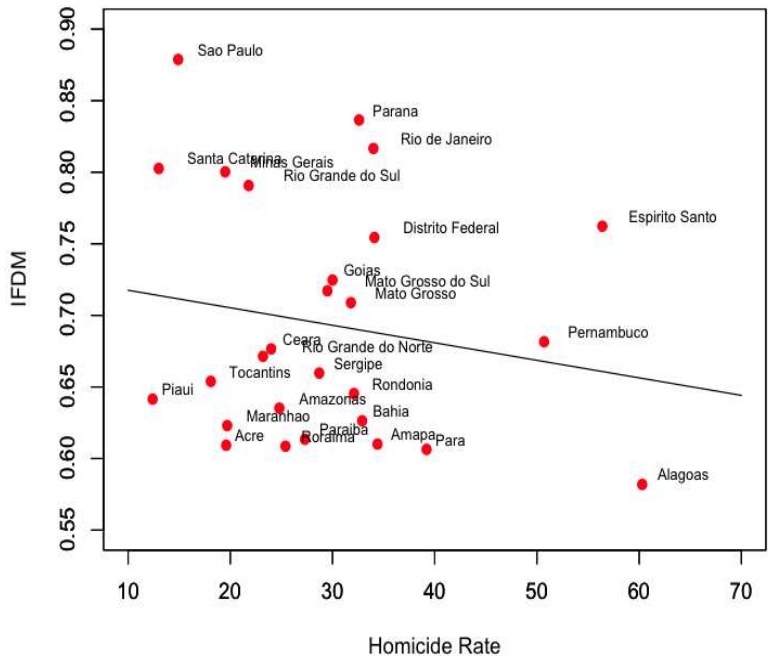}}}
\caption{Scatterplot showing homicide rates and the IFDM index (both corresponding to year 2008). 
The solid line is the best linear fit to the data.}
\label{IFDM_homi_scatterplot}
\end{figure}

Uncovering correlations between crime statistics and other demographic, social, and economic variables is an active focus of 
research aiming at understanding the conditions that foster criminal behavior. For instance, Gould et al. 
found that low wages and unemployment make less-educated men more likely to turn to crime~\cite{goul02}. 
The impact of unemployment on crime has also been studied very recently using country-level statistical data from Europe~\cite{alti11}. Other recent studies on crime-rate regressions are e.g. Refs.~\cite{mood10, cott11, deto11} and references therein. 
Following our case study of state-level homicide rates in Brazil, let us now illustrate how correlations can be put in evidence 
by using diffusion-based cartograms. 

As anticipated above, we used the IFDM index calculated from the so-called ``FIRJAN System"~\cite{firjan}, which measures the combined education, health, and economic development at the municipal level based on official data released by the Brazilian  Ministries of Education, Health, and Labor. Since the IFDM index is a measure of welfare and prosperity, it is reasonable to 
expect states with higher IFDM scores to be less affected by violent crimes, and viceversa. Indeed, the scatter plot in 
Fig.~\ref{IFDM_homi_scatterplot} hints to a plausible correlation between the two statistics, as evidenced by a best linear 
fit with negative slope (shown by the solid line) and a Pearson's correlation coefficient equal to $-0.17$. Notice, however, that 
the p-value for the null correlation hypothesis is 0.385, and therefore, from a strict 
statistical perspective, we cannot rule out the absence of correlations. In the context of this paper, our choice of the IFDM index is only guided by the purpose of illustration of the cartogram method. 

\begin{figure}[t!]
\centerline{{\epsfxsize=3.5in \epsfysize=4.0in \epsfbox{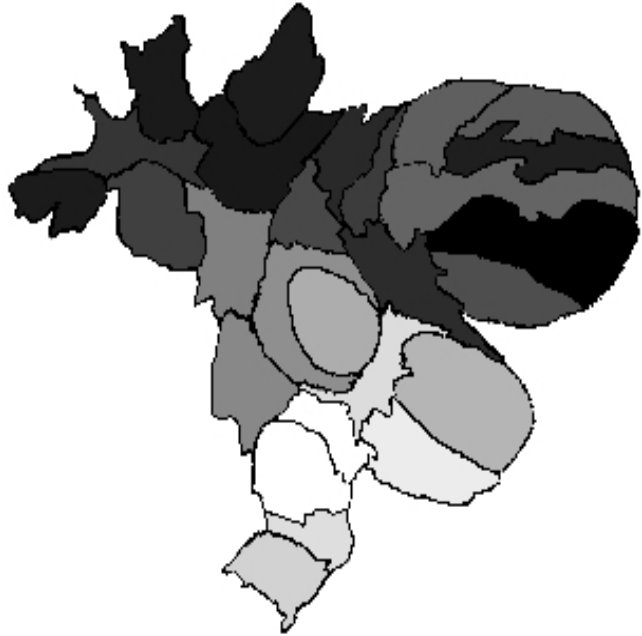}}}
\caption{Cartogram showing areas proportional to each state's homicide rates per 100,000 inhabitants in 2008 
and grayscale shading corresponding to each state's IFDM index (where darker states have lower IFDM scores). 
The cartogram puts in evidence the high geographical correlation of the IFDM index, with a dark fringe crossing the Northern and 
North-Eastern regions, while the states become increasingly lighter towards the South. Moreover, the cartogram captures plausible  
correlations between the IFDM index and crime rates, 
as for instance the large, dark North-Eastern regions (indicating low development and high crime rates) and, 
inversely, the comparatively smaller and lighter Southern states.}
\label{IFDM_homi_cartogram}
\end{figure}

Fig.~\ref{IFDM_homi_cartogram} shows a cartogram that, as Fig.~\ref{absolute_crime_popul}, displays the area of each 
state proportional to each state's homicide rates per 100,000 inhabitants in 2008. The grayscale shading represents 
each state's IFDM index (where darker states have lower IFDM scores). On the one hand, this cartogram puts in evidence the high geographical correlation of the IFDM index, with a dark fringe crossing the Northern and 
North-Eastern regions, while the states become increasingly lighter towards the South. This kind of geographical correlation, 
immediately apparent in the cartogram representation, is more difficult to spot in the scatter plot of Fig.~\ref{IFDM_homi_scatterplot}. On the other hand, moreover, the cartogram conveys plausible  
correlations with crime rates, as for instance the large, dark North-Eastern regions (indicating low development and high crime rates) and, inversely, the comparatively smaller and lighter Southern states. 
      
Finally, another interesting application of the diffusion-based cartogram representation is to capture the dynamical evolution of 
statistical quantities. By using annual 
data on homicide rates from 1998 to 2008, we generated additional frames by linear interpolation 
and created a movie showing substantial changes experienced over time (supplied as Electronic Supplementary Material to this paper). As opposed to analogous movies based on the usual color- or gray-scale representation 
(Fig.~\ref{usual_representation}), cartogram movies display changing features such as shape and size, and therefore are particularly suitable for enhancing the visualization of dynamical observables. This technique can certainly improve 
the communication and outreach of statistical results to both specialized and non-specialized audiences.    

\section{Conclusions and Outlook}
Geography and statistics are ubiquitously intertwined in data collected to assess the levels of health, socioeconomic development, literacy and education, crime, and many other key indicators of the welfare of society. However, georeferenced 
data are most often represented by traditional visualization methods that do a poor job in integrating these two key aspects. 
In this work, we implemented a physics-based visualization approach to integrate geography and statistics into a 
unified framework. 
By trading off geographical accuracy in exchange for additional information, we explored different alternative scenarios 
that emphasize different views on the data. 
We believe that these novel techniques offer new avenues to communicate effectively and reach out to specialized and non-specialized audiences in a wide variety of applied statistics fields, including economics and gross domestic product research, public health and epidemiology, socio-demographics, political science, business and marketing, and many others.     

Although we illustrated the diffusion-based approach with Brazilian georeferenced statistical data at the federative unit level, 
the method can be extended to integrate different geographical and geopolitical scales, from local (e.g. town 
districts) to global (e.g. regions within a country and even relations among countries in the international context). 
Naturally, in order to fully capture the inherent complexities of georeferenced data across multiple dimensions, 
other methodologies are needed to complement the approach presented here. To that end, 
methods based on multi-level Voronoi tesselation~\cite{cand08} and   
multilayered complex network analysis and modeling~\cite{lamb08,cran10}  
offer promising avenues of research.   

Besides the specific aim of improving the visualization of statistics, we hope that, 
in a broader context, this work will serve to illustrate possible cross-disciplinary endeavors addressing key local 
and regional issues. Indeed, we stress the importance of two aspects, namely: (i) recognizing the need for interdisciplinary work in many areas of applied science and the role physicists can play in providing new analysis methods; and (ii) recognizing the need to address key local and regional issues (e.g. within the Latin American context) in those areas where basic and applied research can find a common ground. Hopefully, this paper will contribute to those goals and stimulate further work.

\section*{Acknowledgments}
This work was financially supported by CONICET, UNLP, and ANPCyT (Argentina).

\end{document}